\documentclass[twocolumn,showpacs,preprintnumbers,amsmath,amssymb]{revtex4}

\usepackage{epsfig}
\usepackage{epsf}
\usepackage{graphics}
\usepackage{graphicx}

\begin{document}

\newcommand{\bc}{\begin{center}}
\newcommand{\ec}{\end{center}}
\newcommand{\be}{\begin{equation}}
\newcommand{\ee}{\end{equation}}
\newcommand{\ikl}{\int_k^\Lambda}
\newcommand{\dd}{\frac{d^2k}{(2 \pi)^2}}
\newcommand{\dt}{\frac{d^3k}{(2 \pi)^3}}
\newcommand{\dtp}{\frac{d^3p}{(2 \pi)^3}}
\newcommand{\dq}{\frac{d^4k}{(2 \pi)^4}}
\newcommand{\dn}{\frac{d^nk}{(2 \pi)^n}}
\newcommand{\PLB}{{\it{Phys. Lett. {\bf{B}}}}}
\newcommand{\NPB}{{\it{Nucl. Phys. {\bf{B}}}}}
\newcommand{\PRD}{{\it{Phys. Rev. {\bf{D}}}}}
\newcommand{\AOP}{{\it{Ann. Phys. }}}
\newcommand{\MPL}{{\it{Mod. Phys. Lett. }}}
\newcommand{\del}{\partial}

\title{Quantum Conductance and Electrical Resistivity}
\date{\today}
\author{P. R. Silva}
\email []{prsilva@fisica.ufmg.br}
\author{Marcos Sampaio}
\email []{msampaio@fisica.ufmg.br}
\author{Claudio Nassif}
\email []{cnassif@fisica.ufmg.br}
\author{M. C. Nemes}
\email []{carolina@fisica.ufmg.br}

\affiliation{Federal University of Minas Gerais -
Physics Department - ICEx     \\
P.O. BOX 702, 30.123-970, Belo Horizonte MG - Brazil}

\affiliation{Federal University of Ouro Preto - ICEB\\
Morro do Cruzeiro, 34.500-000, Ouro Preto MG - Brazil
}

\begin{abstract}
\noindent
The Landauer formula for quantum conductance, based on the modern paradigm: ``conduction is transmission",
is generalized to samples of macroscopic size. Two
regimes of electrical conduction, namely diffusive and ballistic ones, are
studied. In the former regime,  Drude's formula for the electrical resistivity
is recovered and it is found a maximum conductivity equal to $(e^2   m c)/(\pi
\hbar^2)$, which is of the same order of magnitude as that of good metals at
room temperature. In the latter, it is obtained in three dimensions a quantum
conductance which is compatible with the one deduced by Sharvin in the ballistic
regime. It is also found in this case an electrical conductivity which depends
on the size of the sample, in agreement with that measured in very pure metals
at the temperature of liquid helium. In two dimensions the result for the
conductance in the ballistic regime is consistent with that used to analyse
quantum point contacts.

\end{abstract}
\pacs{72.10.-d, 72.10.Bg, 72.20.-i, 72.20.Dp, 73.23.Ad, 73.50.Bk }
\keywords{Conductance}
\maketitle
\section{Introduction}

A conventional view of the electrical conductivity attributes
the onset of it to the linear response of the free electrons to the applied
external electric field. This picture is contemplated both in classical
Drude-Sommerfeld-Lorentz and in the quantum mechanical Kubo treatments.
On the other hand a modern view of the electrical conductance was proposed by
Landauer \cite{1}, \cite{2}, which states that conduction is transmission (see
also    van Houten and Beenakker \cite{3}). Both the conventional and modern
views are treated in a paper by Rammer \cite{17}   in which, among other
relevant considerations,   discusses the connection between linear response
formalism and the Landauer approach by expressing the conductance in terms of
scattering properties of the sample.

As pointed out by Batra in \cite{4}, it is well known that for an ideal one
dimensional conductor under ballistic transport conditions, the conductance is
transversely quantized in steps of $ 2 e^2/h $ as the constriction width is
varied. According to Batra \cite{4}, \cite{5}, the finite resistance of a
perfect conductor, which has been previously understood in terms of contact
effects, can also be viewed as having quantum mechanical origin in the
uncertainty principle.

Landauer \cite{1}, \cite{2} obtained a relation of the conductance for a one
dimensional sample connecting two reservoirs at different electrochemical
potentials through ideal conductors. In the absence of dissipative scattering
the conductance $G$ is given by
\be
G = \frac{e^2}{\pi \hbar} T \, ,
\ee
 where $T$ is the transmission probability of the sample, $e$ the quantum
of electrical charge and $\hbar$ the reduced Planck's constant.

In the particular case where we have $N$ perfectly transmitting channels, the
conductance becomes
 \be
G = \frac{e^2}{\pi \hbar} N \, .
\label{eqn2}
\ee
 Experimental verification of equation   (\ref{eqn2}),     comes from the study
of quantum point contacts (QPC) \cite{6}, \cite{7}. In a two dimensional
heterojunction the channel width can be controlled by external applied gate
voltages. The conducting channel works approximately as a waveguide. As it is
widened, the number of transverse eigenstates below the Fermi level increases.
Then conductance steps corresponding to different values of $N$ in equation
(\ref{eqn2}), are observed \cite{2},   \cite{3},  \cite{6},  \cite{7}.

In   this letter we aim    to extend the validity of equation     (\ref{eqn2})
to the case of macroscopic samples in any spatial dimension, once the number $N$
of conducting channels is properly interpreted. We intend to make    connections
of relation (\ref{eqn2}) with the conduction of electricity in good metals at
the room temperature, as well with very high impurity-free samples at very low
temperatures. We will see that in the first case it is possible to deduce the
Drude formula starting from the Landauer relation for te quantum conductance
and also to estimate the ``real" conductivity of a good metal in room
temperature. In the second case we will obtain an electrical conductivity which
depends on the size of the sample. We will also compare these results with those
obtained by Sharvin \cite{8} and by Lifshits and Kaganov \cite{9}, \cite{10}.
Finally the ballistic regime will be treated from the point of view of a drag
force induced by a turbulent flow.

As a starting point of this work let us take a hypercubic sample of a good
conductor of size $L$. By considering that the free electrons which are able to
participate in the electric conduction are those close to the Fermi level with
momentum $p_F$, it is convenient to write the following action in momentum space
\be
A = \int_{(\Delta p)^D} \Big( | \vec{\nabla}_p \psi |^2 - \frac{|\psi|^2}{ 2 m
E} \Big) \, d^D \vec{p}   \, ,
\label{eqn3}
\ee
where         $ \Delta p $    is the wavepacket width around
$\vec{p_F}$  , $m$ is the electron mass,  $E$ the energy and $D$ is the space or
the space-time dimension.

Thompson \cite{11} has introduced  a heuristic method (of the dimensions) as a
tool to deal with the critical behavior of a system undergoing a second order
phase transition \footnote{For applications of Thompson's method see
\cite{aplic}.}. One of the basic hypothesis of Thompson is that each term of the
Landau-Ginzburg-Wilson free energy used to describe the cooperative system is
separately scaled to be of the order of the unity. In what follows  we extend
this framework  to the action given by equation (\ref{eqn3}). Now if each term
of the action (\ref{eqn3}) is separately taken to be of order of the unity, we
may substitute equality between integrals with equality between integrands
\be
 \vec{\nabla}_p \psi = \pm \frac{i}{\sqrt{2 m E}} \psi  \, .
 \label{eqn4}
\ee
A solution of equation (\ref{eqn4}) gives
\be
 \psi = \psi_0 e^{\pm \frac{i}{\sqrt{2 m E}} p} \, .
 \label{eqn5}
\ee
It is possible based on   (\ref{eqn5}) to construct a wavepacket centered on
$p_F$. From the first term in the action (\ref{eqn3}), by using Thompson's
prescription we may also write
\be
 \int_{(\Delta p)^D}  |  \vec{\nabla}_p \psi  |^2  \, d^D \vec{p} \sim
 |\psi|^2_{av} (\Delta p)^{D-2} \sim 1 \, .
\label{eqn6}
\ee
Equation (\ref{eqn6}) implies that
\be
|\psi|^2_{av} \sim    (\Delta p)^{2-D}  \, ,
\label{eqn7}
\ee
 where $``av"$ stands for the averaged quantity. Taking into account the
uncertainty relations, namely $ \Delta p L \sim \hbar$, enables us to write
equation (\ref{eqn7}) as
\be
|\psi |^2_{av} \sim L^{D-2} \, .
\label{eqn8}
\ee
We interpret  $|\psi|^2_{av}$      above as the transmission probability of the
sample and we will consider two possible regimes  of conduction, namely the
classical regime where $D$ stands for the $d$-spatial dimensions and the quantum
regime where $D=d+1$ is the $d$-spatial plus one time dimension. Therefore
combining equations (\ref{eqn2}) and (\ref{eqn8}) we obtain
\be
G = \frac{e^2}{\pi \hbar} \Big( \frac{L}{l_0} \Big)^{D-2}  \, ,
\label{eqn9}
\ee
 where $l_0$ is the size of a typical channel.

In their work about the scaling theory of the localization, Abrahams et al
\cite{12} have defined a generalized dimensionless conductance that they called
``Thouless number" as
\be
g(L) = \frac{G(L)}{e^2/ 2 \hbar} \, .
\label{eqn10}
\ee
 Therefore we can identify $(L/l_0)^{D-2}$ as the ``Thouless number" in the
delocalized diffusive regime.

\section{The Classical Regime}

In the classical regime $D$ coincides with the spatial dimensionality $d$ and
here we are particularly interested in the case $d=3$. Thus from equation
(\ref{eqn9}) we have in three dimensions
\be
G_{D=3} = \Big( \frac{e^2}{\pi \hbar l_0}  \Big) L \, .
\ee
On the other hand, for large $g$, macroscopic transport theory is correct and
gives \cite{12}
\be
G(L) = \sigma L^{d-2} \, ,
\label{eqn12}
\ee
where $\sigma$is the electric conductivity.
Comparing (\ref{eqn12})  with  (\ref{eqn9})  we get, in the three dimensional
case,
\be
\sigma = \frac{e^2}{\pi \hbar} \frac{1}{l_0} \, ,
\label{eqn13}
\ee
where $l_0$ can be evaluated through the following reasoning.  Suppose that we
have $n$ scatterers per unity of volume and let us consider a cylinder shaped
tube with longitudinal size equal to the electron mean free path $\lambda$ and
radius equal to the geometric average of $l_0$ and $\ell_F$, where  $\ell_F =
l_F/(2 \pi)$ is the reduced Fermi wavelength. It must be stressed that $n$ is
numerically equal to the number of electrons per unity of volume, if we consider
that electric conductivity always happens in a regime of charge neutrality. We
write
\be
n \pi l_0 \ell_F \lambda = 1 \, .
\label{eqn14}
\ee
Inserting $l_0$ given by (\ref{eqn14}) into      (\ref{eqn13}) allows us to
obtain
\be
\sigma = \frac{e^2 n \lambda}{m v_F} \, ,
\label{eqn15}
\ee
where $v_F$ is the Fermi velocity of the charge carrier.   Equation
(\ref{eqn15}) is just the Drude formula for the electrical conductivity, but
here it was deduced starting from an expression describing quantum conductance.

It is also worthwhile noticing that if $l_0$ is considered as the width of the
transmission channel then there should be a lower bound for it, namely the
reduced Compton wavelength of the electron. Inserting $l_0 = \hbar/(mc)$ in
(\ref{eqn13})  yields
\be
\sigma_{max} = \frac{e^2 m c }{\pi \hbar^2} \, ,
\label{eqn16}
\ee
where $\sigma_{max}$ means the maximum conductivity in the classical (diffusive)
regime. A  numerical evaluation of (\ref{eqn16}) gives $\sigma_{max} \sim 10^8
(\Omega . m)^{-1}$. Indeed electrical conductivities of this order of magnitude
are typical of those measured in good metals at room temperature.

Another evaluation of the electrical resistivity of metals was accomplished by
Lifshits and Kaganov \cite{9}, \cite{10} (see also Brandt and
Chudinov \cite{15}). The result obtained by Lifshits and Kaganov (LK) is
\be
\sigma_{LK} = \frac{2}{3} e^2 \frac{S_F \lambda}{(2 \pi \hbar)^3} = \frac{4}{3}
\frac{e^2}{\hbar} \frac{\lambda}{l_F^2} \, ,
\label{eqn24}
\ee
where $S_F = 4 \pi p_F^2$ and $\lambda$ is the electron mean free path.
Comparing (\ref{eqn24}) with (\ref{eqn13}) yields
\be
\lambda = \frac{3}{4 \pi} \frac{l_F^2}{l_0} \, .
\label{eqn25}
\ee

\section{The Quantum Regime}

In the quantum regime we will take $D$ to be equal to $d+1$ ($d$ spatial plus
one time dimension). In this case equation (\ref{eqn9}) becomes
\be
G = \frac{e^2}{\pi \hbar} \Big(\frac{L}{l_0}\Big)^{d-1}  \, .
\label{eqn17}
\ee
It would be interesting to make a detailed  investigation of (\ref{eqn17}) in
the special cases of two and three  spatial dimensions ($d=2$ and $d=3$).
If we make the natural choice of identifying in (\ref{eqn17}) $l_0$
with $l_F$  (the Fermi wavelength), we get in three dimensions
\be
G_{d=3} = \frac{e^2}{\pi \hbar} \Big( \frac{L}{l_F} \Big)^2 \, .
\label{eqn18}
\ee
By comparing (\ref{eqn18}) with relation (\ref{eqn12}) in the case $d=3$,
enables us to write
\be
\sigma_{d=3} = \frac{e^2}{\pi \hbar}  \frac{L}{l_F^2}
\ee
We observe that in the case of very pure metallic crystals at the liquid helium
or lower temperatures, the electrical conductivity depends on the size of the
sample. According to C. Kittel \cite{13}, mean free paths as long as $10$ cm
have been observed in very pure metals in the liquid helium temperature range.
It seems that in this case, the mean free path is ultimately determined by the
size of the sample. Taking $L = 10$ cm and $l_F$ as the Fermi wavelength
of the copper, we get a electrical conductivity of the order of magnitude
$10^{14} (\Omega . m)^{-1}$ .

A good description of the electrical conduction for very pure metals at low
tempeartures seems to be the ballistic transport treatment introduced by Sharvin
\cite{8} (see also Brandbyge et al \cite{14}). We have
\be
G_{Sharvin} = \frac{e^2 L \nu}{ p_F } \, ,
\label{eqn20}
\ee
where $p_F$ is the Fermi momentum and $\nu$ is the free electron density. One
electron which has an uncertainty in momentum equal to $p_F$ has an uncertainty
in position equal to $l_F = \hbar/p_F$ . Therefore by considering spin
degeneracy we can write
\be
\nu = \frac{2}{l_F^3} \, .
\label{eqn21}
\ee
Inserting (\ref{eqn21}) into (\ref{eqn20}) leads to
\be
G_{Sharvin} = \frac{e^2 }{\pi \hbar}\Big(   \frac{L}{l_F}     \Big)^2 \,  ,
\label{eqn22}
\ee
which coincides with our equation (\ref{eqn18}) (see also Brandbyge et al
\cite{14}).

Now let us analyse the two dimensional case in the ballistic regime. From
equation (\ref{eqn17}) we have
\be
G_{d=2} = \frac{e^2}{\pi \hbar} \Big( \frac{L}{l_F} \Big) \, .
\label{eqn23}
\ee
When the measurement is done at very low temperatures in a very pure two
dimensional macroscopic sample, the quantization of the conductance cannot be
detected since we have in this case $L \gg l_F$. However, measurements in
Ga As-Al Ga As heterojunctions show that each new channel of transmission is
activated only when the width $w$ increases as $\Delta w \equiv \Delta L = l_F$.
This characterizes the onset of the quantum conductance as it can be seen in the
work by van Houten and Beenakker \cite{3} (please see also \cite{6}
and \cite{7}) .

If we consider as before that the size of the transmission channel is limited by
the electron reduced Compton wavelength, we will have in the diffusive regime an
upper bound to the electron mean free path given by
\be
\lambda_{max} = \frac{3}{4 \pi}\frac{m c \, l_F^2}{\hbar} \, .
\label{eqn26}
\ee
Order of magnitude estimates of (\ref{eqn26}) results in
$\lambda_{max} \sim 10^4 \AA$ for charge carriers in semiconductors and
$\lambda_{max} \sim 10^2 \AA$ for electrons in metals.

\section{An Alternative View of the Ballistic Regime}

A macroscopic body moving at speeds high enough so that the flow of air behind
it is turbulent, is subject to a drag force $D$ given by (see for instance
Halliday and Resnick \cite{16})
\be
D = \frac{1}{2} C \rho A v^2 \, .
\label{eqn27}
\ee
Here $A$ is the effective cross-sectional area of the body, $\rho$ is the
density of the fluid, $v$ is the speed of the body and $C$ a  dimensionless
coefficient.

We think that in the ballistic regime the electrical resistance of a good
conductor can be represented by the collisions of the charge carriers with walls
which have the size of the sample, so that the area of the walls are $L^2$. We
observe that the ultimate wall is that which separates the sample from the
surrounding dielectric medium. Changing the reference frame, we can imagine a
wall being  draged by the ``fluid" composed by the free electron gas. With these
ideas in mind, we can suppose that the dissipation $P$ is given by
\be
P = D v_F =\frac{1}{2} C \rho A v_F^3 \, ,
\label{eqn28}
\ee
where $\rho = n m$, $n$ and $m$ being respectively the number density and mass
of charge carriers. Now impose the equality between the drag power (equation
(\ref{eqn28}) and the power dissipation due to Joule effect, namely,
\be
P_J = G V^2 \, ,
\label{eqn29}
\ee
where $G$ is the electrical conductance   and $V$ is the applied potential
difference. Using $n$ given by equation (\ref{eqn14}) and considering the
equality between the powers given by (\ref{eqn28}) and (\ref{eqn29}), we can
write \be
\frac{C m^2 v_F^4 A}{ h l_0 \lambda} = G V^2 \, .
\label{eqn30}
\ee
Finally using that
\be
\frac{1}{2} m v_F^2 = e V \, ,
\ee
we obtain
\be
G = \frac{2 e^2}{h} \Big( \frac{L^2}{l_0 \lambda} \Big) \, ,
\label{eqn32}
\ee
where we have fixed $C=1/2$ and $A=L^2$.

Sharvin's result can be recovered if we put
\be
l_0 \lambda = l_F^2 \, .
\ee
Although (\ref{eqn32}) has been deduced specifically for the three dimensional
case, it can be easily extended to other dimensions. In two dimensions we could
think in terms of a ``line wall" of size $L$ whereas the drag power can be
written as (compare with (\ref{eqn28}))
\be
P_{d=2} = \frac{1}{2} C \rho_s L v_F^3 \, ,
\label{eqn34}
\ee
where
\be
\rho_s = n_s m \, ,
\ee
$n_s$  being the surface density of charge carriers. Working in an analogous way
we have done before we get, after comparing (\ref{eqn34}) with   (\ref{eqn29}),
\be
G_{d=2} = \Big( \frac{2 e^2}{h}\Big)  \Big(  \frac{L}{l_0} \Big)  \, ,
\ee
where we have considered $n_s = \frac{2}{l_F l_0} $ and $C=1/2$.

Again  we recover (\ref{eqn23}) if we take  $l_0 = l_F$.

All these considerations permit us, after taking into account the drag force, to
generalize the formula for the electrical conductance in the ballistic regime,
which reproduces equation (\ref{eqn17}) of this work.

\section{Concluding Remarks}

It is interesting to remember that in the paper on the scaling
theory of localization, Abrahams et al \cite{12} have introduced the ``Thouless
number" (equation (\ref{eqn10}) of this paper). According to the present work it
is possible to distinguish two classes of such quantity, one of them being
$g_{diff}=(L/l_0)^{d-2}$, referring to the diffusive regime and the other
$g_{ball}=(L/l_0)^{d-1}$ referring to the ballistic regime of the electrical
conduction.

Finally it is worth to mention that both Landauer transport theory and Kubo's
formula were used to compute DC conductance in a impurity system in a recent
work by  Castro-Alvaredo and Fring \cite{18}. They found an identical
plateau structure for the DC conductance in the ultraviolet limit, displaying
the agreement between the two approaches.


\begin{thebibliography}{88}

\bibitem{1} R. Landauer, IBM J. Res. Dev.  {\bf{1}}, 223 (1957).
\bibitem{2} Y. Imry and R. Landauer, Rev. Mod. Phys.  {\bf{71}}, S 306 Centenary
(1999).
\bibitem{3} H. van Houten and C. Beenakker, Phys. Today  {\bf{49}}, 22
(July 1996).
\bibitem{17} J. Rammer, Rev. Mod. Phys.  {\bf{63}}, 781 (1991).
\bibitem{4} I. P. Batra, Surf. Sci.  {\bf{395}}, 43 (1998) .
\bibitem{aplic} C. Nassif, P. R. Silva, Mod. Phys. Lett. B {\bf{13}}, 829
(1999);  Mod. Phys. Lett. B {\bf{15}}, 33 (2001); Mod. Phys. Lett. B {\bf{15}},
1205 (2001); Mod. Phys. Lett. B {\bf{16}}, 601 (2002); P. R. Silva, Phys. Stat.
Sol. (b)  {\bf{179}}, K5 (1993); Phys. Stat. Sol. (b)  {\bf{165}}, K79 (1991);
Phys. Stat. Sol. (b)  {\bf{174}}, 497 (1992); Phys. Stat. Sol. (b)  {\bf{179}},
K99 (1993); P. R. Silva and V. B. Kokshenev, Braz. J. Phys.  {\bf{30}}, 783
(2000).
\bibitem{5} I. P. Batra, Sol. State Comm.  {\bf{124}}, 463 (2002).
\bibitem{6} B. J. van Wees, H. van Houten, C. W. J. Beenakker, J. G. Williamson,
L. P. Kouwenhoven, D. van der Marel, C. T. Foxon, Phys. Rev. Lett.  {\bf{60}},
848 (1988); Phys. Rev. B  {\bf{43}}, 12431 (1991).
\bibitem{7} J. M. Krans, J. M. van Ruitenbeek, V. V. Fisun, I. K. Yanson, L. J.
de Jongh, Nature  {\bf{375}}, 767 (1995).
\bibitem{8} Yu. V. Sharvin, Sov. Phys. JETP {\bf{ 21}}, 655 (1965).
\bibitem{9} I. M. Lifshits and M. I. Kaganov, Uspekhi Fiz. Nauk  {\bf{69}}, 419
(1969); Uspekhi Fiz. Nauk  {\bf{78}}, 411 (1962).
\bibitem{10} I. M. Lifshits, Sov. Phys. JETP  {\bf{38}}, 1569 (1960).
\bibitem{11} C. J. Thompson, J. Phys. A  {\bf{9}}, L25 (1976).
\bibitem{12} E. Abrahams, P. W. Anderson, D. C. Licciardello and T. V. Ramakrishnan,
Phys. Rev. Lett.  {\bf{42}}, 673 (1979).
\bibitem{13} C. Kittel, Introduction to Solid State Physics, Wiley, New York,
1976, p. 169.
\bibitem{14} M. Brandbyge, J. Schi$\phi$tz, M. R. S$\phi$rensen, P. Stoltze, K. W.
Jacobsen, J. K. N$\phi$rskov, L. Olesen, E. Laegsgaard, I. Stensgaard and F.
Besenbacher, Phys. Rev. B  {\bf{52}}, 8499 (1995).
\bibitem{15} N. B. Brandt and S. M. Chudinov, Electronic Structure of Metals, Mir
Publishers, Moscow, 1975.
\bibitem{16} D. Halliday, R. Resnick, Fundamentals of Physics, Third Edition
Extended, Wiley, New York, 1988, p. 109.
\bibitem{18} O. Castro-Alvaredo, A. Fring, Nucl. Phys. {\bf{B649}}, 449 (2003).












\end{thebibliography}
\end{document}